\documentclass[
]{ceurart}

\usepackage{libertinus}

\begin{document}

\copyrightyear{2022}
\copyrightclause{Copyright for this paper by its authors.
  Use permitted under Creative Commons License Attribution 4.0
  International (CC BY 4.0).}

\conference{CHR 2022: Computational Humanities Research Conference, December 12 -- 14,
  2022, Antwerp, Belgium}

\title{Lost Manuscripts and Extinct Texts:\\ A Dynamic Model of Cultural Transmission}


\author[1,2]{Jean-Baptiste Camps}[%
email=jean-baptiste.camps@chartes.psl.eu,
url=https://github.com/Jean-Baptiste-Camps/,
]
\cormark[1]
\fnmark[1]
\address[1]{Venice Center for Digital and Public Humanities, Univ. Ca'Foscari, Fondamenta Malcanton 5449, Venezia, 30123, Italy}
\address[2]{Centre Jean-Mabillon, École nationale des chartes, Paris Sciences \& Lettres, 65 rue de Richelieu, Paris, 75002, France}
\author[3,4]{Julien Randon-Furling}[%
orcid=0000-0003-0385-7037,
email=julien.randon-furling@cantab.net,
]
\cormark[1]
\fnmark[1]
\address[3]{Université Paris-Saclay, ENS Paris Saclay, CNRS, SSA, INSERM, Centre
Borelli, F-91190, Gif-sur-Yvette, France}
\address[4]{FP2M (FR2036), Université Paris-1 Panthéon-Sorbonne, CNRS, SAMM, F-75013 Paris, France}

\cortext[1]{Corresponding author.}
\fntext[1]{These authors contributed equally.}
\begin{abstract}
How did written works evolve, disappear or survive down through the ages? In this paper, we propose a unified, formal framework for two fundamental questions in the study of the transmission of texts: how much was lost or preserved from all works of the past, and why do their genealogies (their ``phylogenetic trees'') present the very peculiar shapes that we observe or, more precisely, reconstruct?
We argue here that these questions share similarities to those encountered in evolutionary biology, and can be described in terms of ``genetic'' drift and ``natural'' selection. 
Through agent-based models, we show that such properties as have been observed by philologists since the 1800s can be simulated, and confronted to data gathered for ancient and medieval texts across Europe, in order to obtain plausible estimations of the number of works and manuscripts that existed and were lost.
\end{abstract}


\maketitle

\section{Introduction}

How much do we preserve of the written knowledge, science \cite{cisne_how_2005} and culture of the past? And how representative is what we know compared to what existed? Such fundamental questions depend on the process through which texts were distributed materially. 

Before the advent of the printing press, written texts were circulated in manuscript form. 
In order to make the text available, the author would dictate it to a secretary, or write a draft on wax tablets, papyrus, parchment or, eventually, paper, and this original, authorial, manuscript would then have to be copied manually by a scribe in the form of a new manuscript, and then circulated. Copies could then be used to create more manuscripts, again by manual copying, perhaps by other scribes in other regions at a later date. During this process, successive modifications were introduced in the text, either by error or intentionally, to make the text more suited to its intended audience.  
These alterations in the written sequence that forms the text could then be transmitted to its ``descendants'' by a given manuscript. 
Wear and tear, accidents, fashions caused the destruction of some manuscripts, while others enjoyed the long life of library preservation. In the end, knowledge was lost and some texts went extinct, while other texts gained traction or were eventually preserved for future generations.

\begin{figure}[phtb]
    \centering %
    \includegraphics[width=0.4\textwidth]{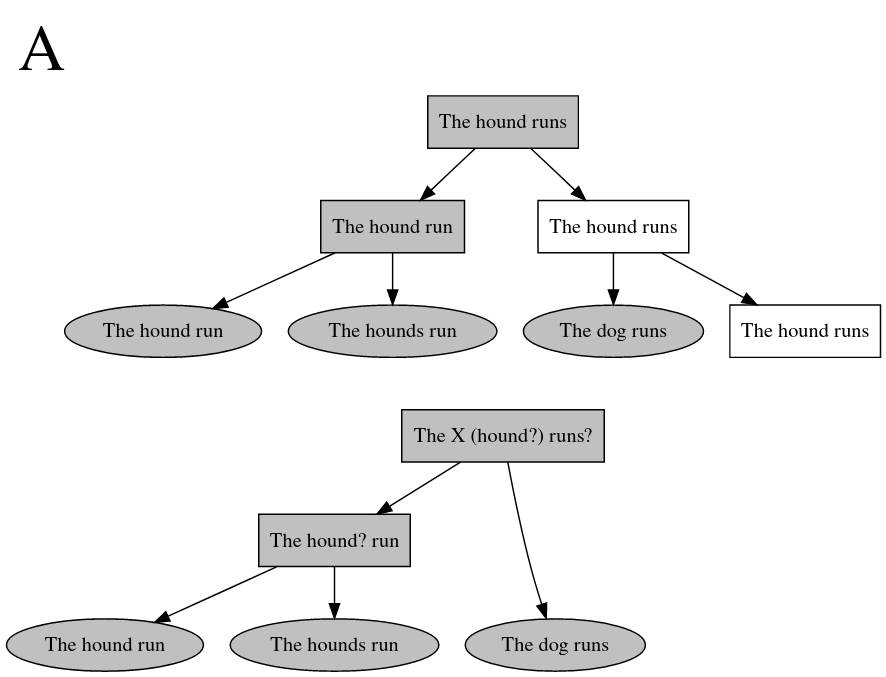}
    \includegraphics[width=0.35\textwidth]{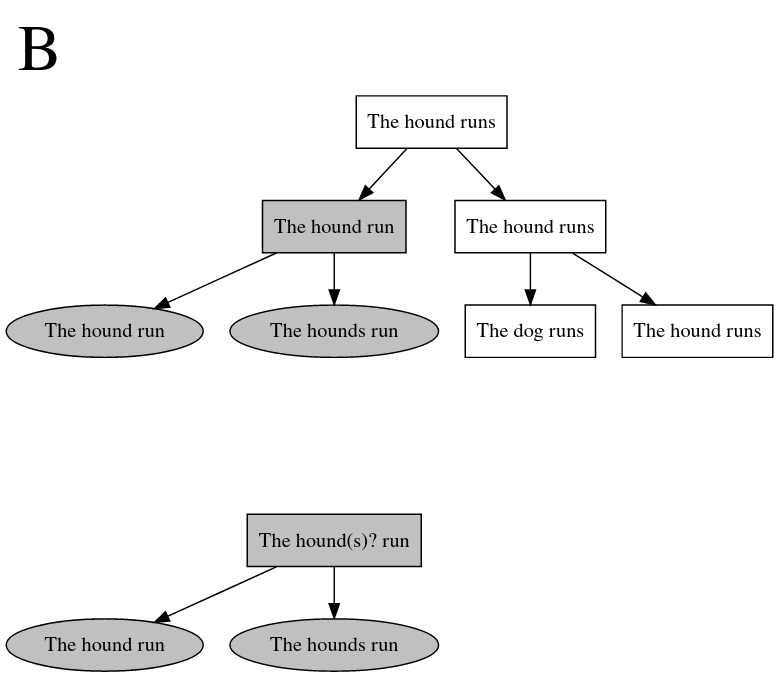}
    \includegraphics[width=0.32\textwidth]{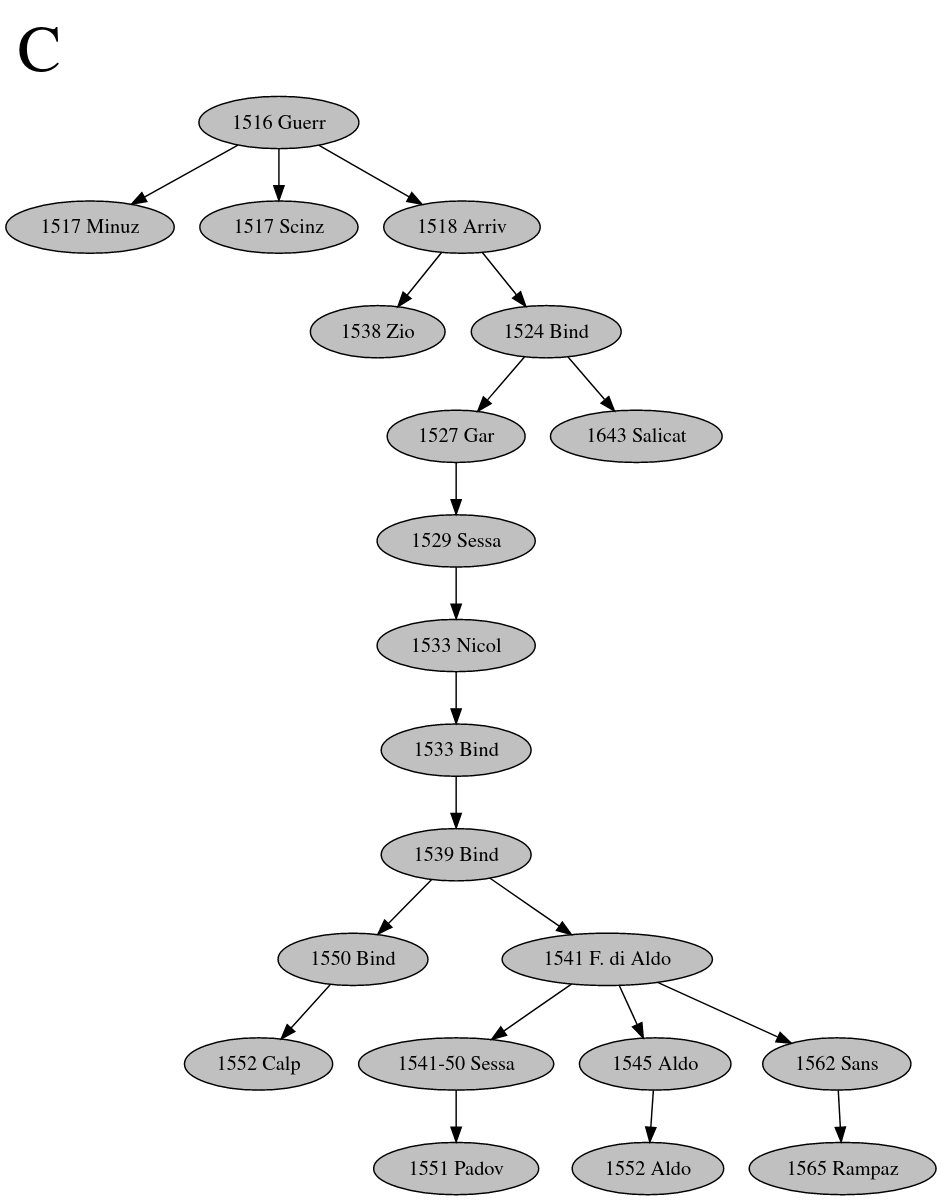}
    \includegraphics[width=0.45\textwidth]{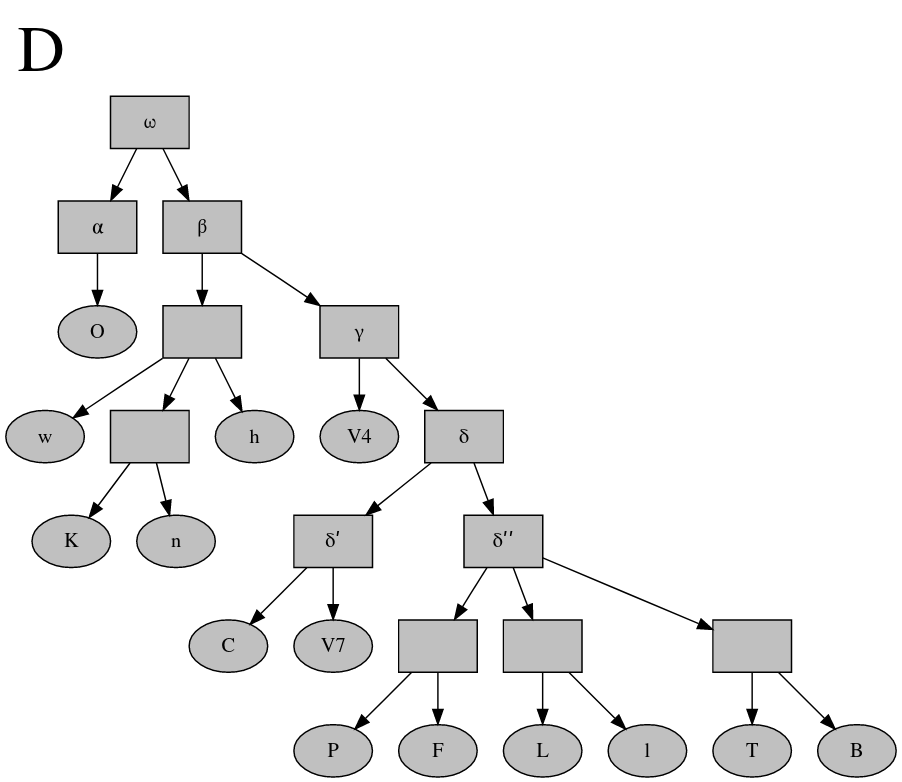}
    \includegraphics[width=0.45\textwidth]{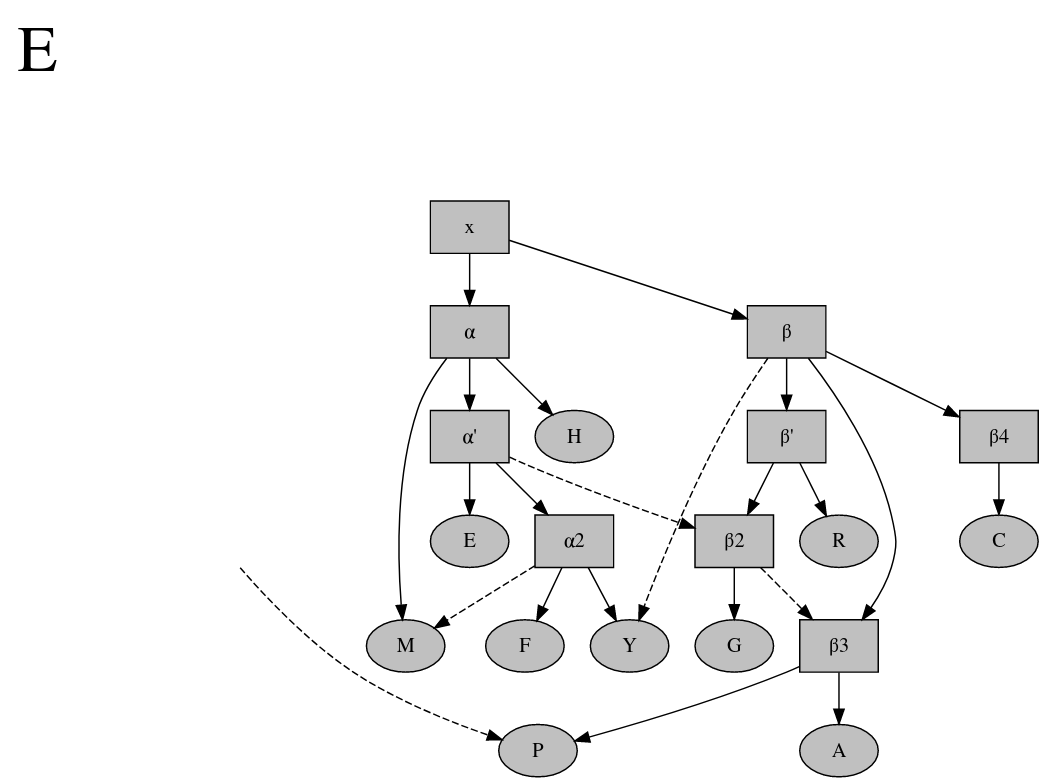}
    \includegraphics[width=0.45\textwidth]{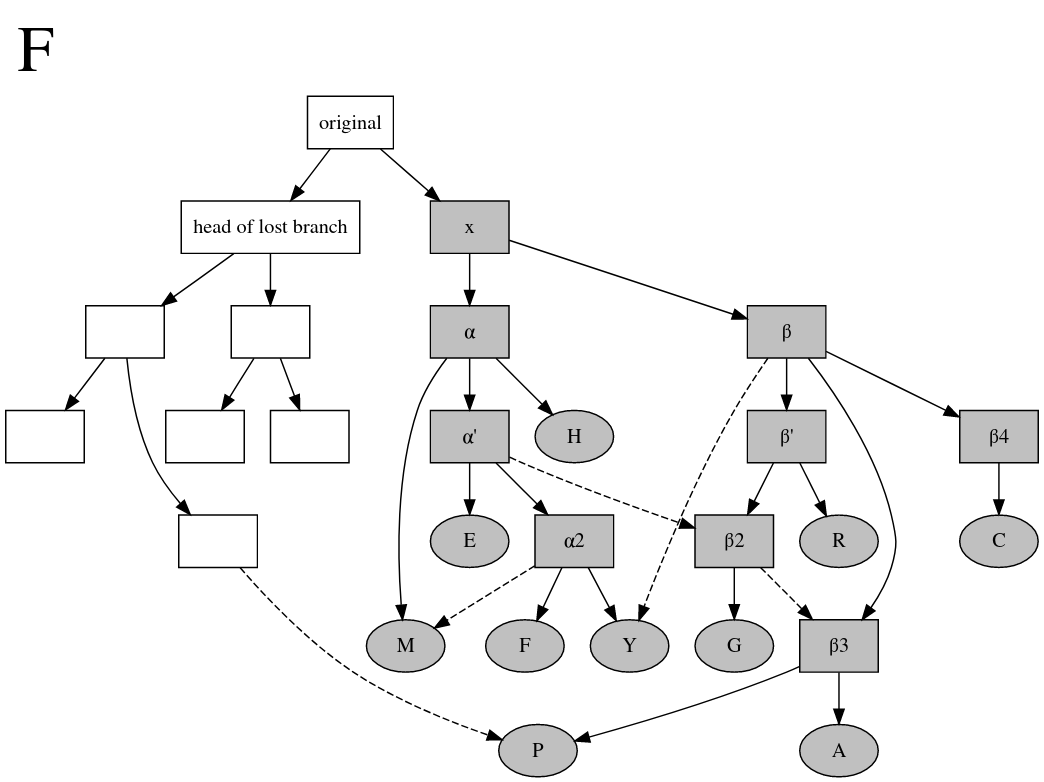}
    \caption{%
    \textbf{From the real tree to the reconstruction}.
    \textbf{A} an artificial example of the 
    transmission of a very short text and its plausible 
    reconstruction, where circles depict extant manuscripts,  rectangles lost ones), and gray, nodes that are recoverable when reconstructing the tree;
    \textbf{B} another artificial example, where the distribution of lost manuscripts causes the loss of a whole branch, and where, consequently, the last common ancestor is not the original (root), but a lower witness (archetype);
    \textbf{C} the observed phylogeny of the successive editions of Fortunio, \textit{Regole grammaticali della volgar lingua} (1516) \cite{fortunio_regole_2001,guidi2004sugli};
    \textbf{D} the reconstructed phylogeny (\textit{stemma}) of the Old French \textit{Song of Roland} \cite{segre_chanson_1971};
    \textbf{E} the reconstructed phylogeny of the Anglo-French \textit{Guy de Warewic} (Ewert, 1932), showing many cases of lateral transmission (``contamination'') in dashed lines, and even one instance of lateral transmission from outside the tree; 
    \textbf{F} the same tree, but including (box shape) the outline of a lost branch (cf.~\cite{hoenen_stemma_2020}).
    }
    \label{fig:fig1} 
\end{figure}

In textual studies and philology, since the development of the ``common errors'' methods from the 19th century onward \cite{timpanaro_genesi_2003,haugen_2_2020}, the analysis of the alterations to the text allows the philologists to reconstruct the relations between the surviving copies (witnesses) of a given work, and to represent them as a tree-like graph, called a \textit{stemma codicum} %
(fig.~\ref{fig:fig1}). 
This reconstructive process is not entirely different from the methods used by biologists to reconstruct the links between existing or extinct species, based on their shared characteristics that are supposed to derive from a common ancestry, and to represent it as a phylogenetic tree. Methods from this biological subfield, known as cladistics,  have even sometimes been directly applied to texts, with controversial results \cite{barbrook_phylogeny_1998,howe_manuscript_2001,spencer_phylogenetics_2004,mace_evolution_2006,hoenen_history_2020}. 
These trees represent the result of an enquiry into the relationships between the surviving witnesses, together with the hypothetical lost nodes that can be deduced from them. To arrive at it, researchers have to study the variants ---~more precisely, common innovations or errors (i.e., mutations)~--- observed in the text of the surviving witnesses (fig.~1, A). 
The reconstructed tree will show only what can be deduced from surviving witnesses: the witnesses themselves, and as many hypothetical nodes as are needed to explain their relationships (fig.~1, B).
While it may be the case with more recent works  that all nodes are known and the graph represents the full transmission of the text ---~for instance with the genealogy of printed editions (fig.~1, C)~---, most of the time the tree represents only a (potentially very small) subset of what existed (fig.~1, D).

A long standing observation, not yet fully understood, about the structure of the reconstructed trees was made by the French philologist Joseph Bédier, 
almost a century ago, in 1928 \cite{bedier_tradition_1928}. 
He observed that most trees reconstructed by philologists show a root bifurcation (a root with outdegree 2): in most reconstructions, 
the original (or archetype) had two, and only two, direct descendants, preventing an accurate reconstruction of the original text by majority principle  (e.g., two witnesses vs one). 
Instead of searching an explanation for such data in the dynamics of text transmission, he interpreted this ``forest of bifid trees'' as the result of 
a methodological flaw or an unconscious bias, spurring a century long debate and causing a long lasting methodological schism inside textual scholarship that remains to be fully resolved~\cite{baker_ombre_2018,duval_tradition_2021}.


Bedier's initial observation has been replicated, with estimates of the proportion of root bifurcation varying, from Bédier's 95.5\%, to somewhat lower estimates ranging from 70\% to  83\% 
\cite{shepard_recent_1930,castellani_bedier_1957,haugen_silva_2015}. 
Yet, some have argued that the prevalence of root bifurcation could be an explainable feature of manuscript transmission of texts as it reached us.
Tentative explanations include combinatorial 
estimations of the proportion of root bifurcation for a given number of witnesses, under the assumption that all configurations are equally likely~\cite{maas_leitfehler_1937,castellani_bedier_1957,Hoenen17}, 
or consider the effects of decimation (i.e., manuscript loss) \cite{greg_recent_1931}, for instance by applying a uniform loss probability to static preexisting trees \cite{guidi2004sugli} or by calculating a node specific loss probability to simulated trees \cite{Hoenen16}. 
Even if it generated little follow-ups, there have also been rare attempts of using birth and death process for exploring the dynamics of manuscript transmission
\cite{weitzman_computer_1982,weitzman_evolution_1987}.

The abundance of roots --~and more generally nodes \cite{haugen_silva_2015}~-- with out-degree 2, is not the only property that can be observed in many stemmas. The asymmetry between branches (cf. fig.~1, C and D), 
the presence or not of lateral transmission (generally called ``contamination''; fig. 1, E) are other properties worthy of investigations, as well as those that can indicate that the tree made from extant witnesses represents only a small portion of the original tradition (i.e., lateral transmission from outside the tree; root identifiable not with the original but with a later manuscript; fig. 1, F). 
It is reasonable to assume that some of these properties reflect the dynamics of manuscript transmission, while others keep trace of  important destruction, decimating manuscripts and removing even full branches. For the texts, this can be seen as an evolutionary process, where two antagonist tendencies are at work: the apparition of textual variants in individuals, causing the increase of diversity in the tradition, and the extinction of full branches, causing some variants to prevail upon others and so  reducing diversity.



Here again, these observations can be put in perspective with problems occurring in evolutionary biology, where, too, two antagonists tendencies are at work, mutation and fixation, either by drift or natural selection, 
in a context where processes of speciation and extinction are strongly linked and 
where extant species represent only a small subset of the species that have existed  \cite{yessoufou_reconsidering_2016}.
In both cases, survival might be the exception and extinction the rule, be it by ``bad genes or bad luck'', a process that can be seen in terms of  gambler's paradox \cite{raup1992extinction, canettieri_philology_2008}.

Basic processes of reproduction and destruction create complex shapes in the trees,
from which one might want to deduce whether they can be fully explained by random process akin to genetic drift, or if differences in selective values are to be suspected. In other terms, going back to textual traditions, if cultural context, through literary taste, canon or fashion for instance, creates  a form of evolutionary pressure on textual traditions.
Any insights gained on this question would have an  applicability beyond the question of the transmission of antique and medieval texts, because it seems that similar dynamics are at work in the diffusion of content in other medium, including print \cite{guidi2004sugli} or
even the web \cite{barabasi_emergence_1999},  and have been observed in areas such as 
the cognitive evolution of scientific fields and the dynamics of scientific memes  \cite{bentley_random_2008,chavalarias_phylomemetic_2013}.

\begin{figure}[phtb]
    \centering
    \includegraphics[width=0.8\textwidth]{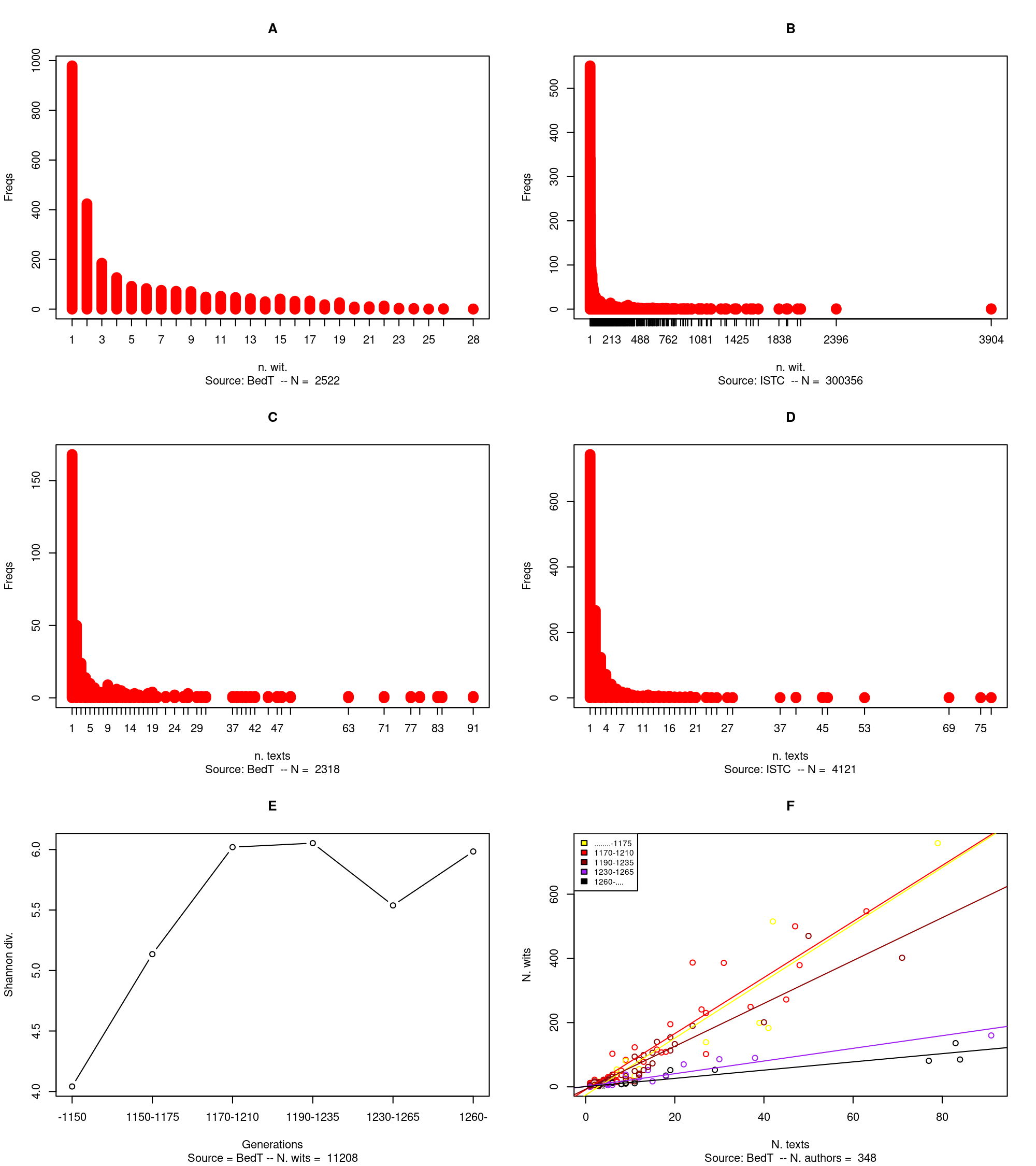}
    \caption{\textbf{A Pareto-like world with diminishing diversity}. 
    \textbf{A} distribution of witnesses per troubadour text; 
    \textbf{B} distribution of the extant copies per work for incunabula printed in Italy;
    \textbf{C} distribution of texts per author for troubadours and  
    \textbf{D} incunabula;
    \textbf{E} Shannon diversity, Generations as sites, texts as species, witnesses as individuals;
    \textbf{F} number of texts and witnesses per author. %
    } 
    \label{fig:pareto}
\end{figure}

Data on cases as different as the songs of the Medieval Occitan troubadours from southern France or the incunabula editions printed in Renaissance Italy outline the same Pareto-like world, where a large number of texts are kept only in a single or handful of documents, while a limited number of ``successful'' texts are kept in a large number of copies (fig.~\ref{fig:pareto}, \textbf{A} and \textbf{B}), where most authors are known only for one or two texts, while a very limited number of writers can have dozens of texts preserved (fig.~\ref{fig:pareto}, \textbf{C} and \textbf{D})… Such a process is also apparent in the constitution of a literary canon of a limited number of authors and texts. This `canonialisation' can be seen has a progressive loss of diversity, where an ever shrinking number of authors and texts take on an ever growing share of the circulated documents (fig.~\ref{fig:pareto}, \textbf{E} and \textbf{F}). But is this due to chance or to a selective process ?

In fact, some properties as were just mentioned for textual traditions have some pendants in evolutionary models, 
concerning for instance the very unequal distribution of descendants  \cite{cosette_single_2015}, 
the varying patterns of biodiversity varying in time and space, studied in macroecology and biogeography 
\cite{cabral_mechanistic_2017,rangel_modeling_2018} and the dynamics of speciation and extinction  
that manifest themselves in the shape of phylogenies and the loss of branches from the tree of life 
\cite{mace_preserving_2003,yessoufou_reconsidering_2016}. 
For this reason, there are inspiration and resources to be found in the study of mathematical properties of evolutionary trees, regarding the establishment of a null-model~\cite{bienvenu_split-and-drift_2019}. 



Can we gain some insight on what existed, what was lost, and the driving forces between extinction or survival of texts ---~by drift or selection? In order to do so, we need a better understanding of the dynamical processes of manuscript transmission.




\section{An agent-based model of a stochastic process}

In this paper we present a selection of the first results obtained with a stochastic model for the transmission of manuscripts in the Middle Ages. Following Weitzman \cite{weitzman_computer_1982,weitzman_evolution_1987}, we use so-called \textit{birth-and-death} processes. These are random processes introduced in probability theory to describe, among other things, simple population dynamics and genealogies. For the simplest versions of theses processes, it is possible to derive analytically (i.e.~with mathematical formulae) certain quantities of interest: the expected number of individuals (here for us, manuscripts) still present at a time $t$, the extinction probability, the survival probability,... But also, quantities of particular interest in the context of manuscript genealogies: the probability that the latest common ancestor (lca) be an archetype rather than the original, or the probability of root bifidity for the reconstructed stemma. Here we favour a numerical approach, that allows us to explore more complicated variants of a birth-and-death process through an agent-based computer simulation. The agents correspond to manuscripts, and during each time step of the simulation each agent has a probability $\lambda$ of being copied (once) and a probability $\mu$ of disappearing. Following Cisne, we limit the increase rate of the population by letting the birth rate $\lambda$ depend on the size of the population at the previous step $k_{t-1}$:
\begin{equation}
   \lambda_{t} = \lambda\; \frac{K}{k_{t-1}} ,
\end{equation}
where $K$ is a theoretical upper bound on the maximum possible size of the manuscript population extant at any given time during the period under consideration.

\section{Phase diagrams obtained through computer simulations}

We ran agent-based simulations of a Cisne-type tradition with a total time frame of $500$~pseudo-years ---~that is, a time-step in the simulation corresponds roughly to $3$~months. This was derived using an estimate for the time taken to produce an average $200$~page manuscript (though the speed of scribes is known to vary a lot, from $1$ to $10$~leaves a day)~\cite{overgaauw_fast_1995}. We chose $K=100~000=10^5$, as an order of magnitude (rather than $10^4$ or $10^6$) for the total number of manuscripts in a given tradition that could be extant simultaneously at any one time\footnote{%
    This order of magnitude is a very rough estimate based on human population and our own assumptions: the medieval population of countries such as France or Italy was in the $10^7$ range; we estimate that the maximum saturation of this market for a given book would be reached if around 1\% of the population were to own a copy. 
}. 

\begin{figure}[phtb]
    \centering
    \includegraphics[width=\textwidth]{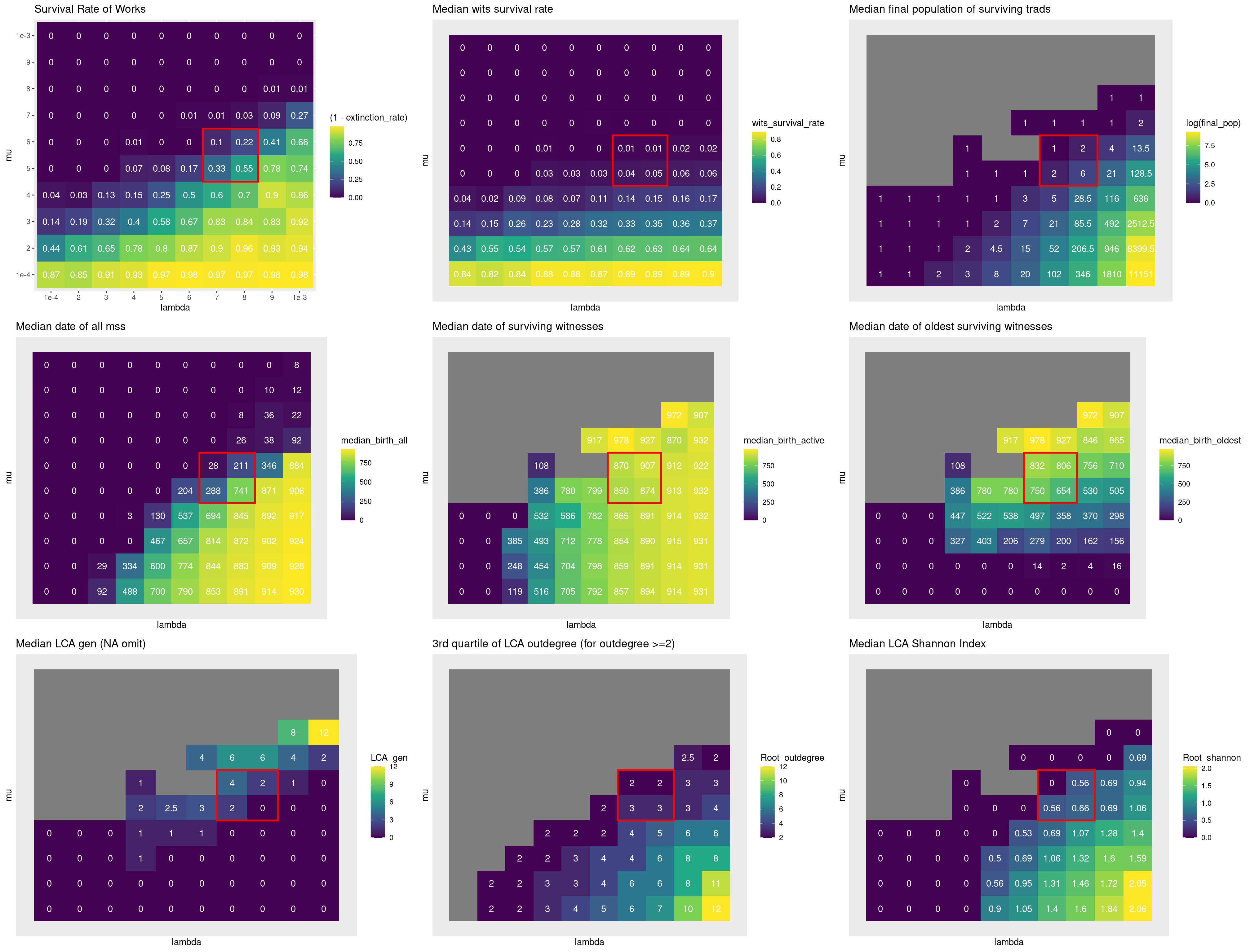}
    \caption{Heat maps (phase diagrams) for the simulation of Cisne-like models; first row contains population size and survival properties, namely 
    \textbf{A} Survival rate of traditions (i.e., trees);
    \textbf{B} survival rate of manuscripts (i.e., nodes);
    \textbf{C} the median final population of surviving witnesses for traditions with at least one;
    second row contains data about the age of manuscripts in the simulation, for 
    \textbf{D} all manuscripts ever created;
    \textbf{E} extant ones;
    as well as \textbf{F} the date of the oldest extant manuscript;
    the third row concerns structural properties of the trees themselves, 
    \textbf{F} the median distance between the lowest common ancestor of the surviving manuscripts and the actual root of the original tree;
    \textbf{G} the third quartile of the LCA (archetype) out-degree and
    \textbf{H} and the median Shannon index for the families (the main branches stemming from the LCA). 
    For each pair of parameter values between 0.0001 and 0.001, 100 simulations were run for 1000 active and 1000 inactive steps. Grey areas correspond to irrelevant values of the parameters, or unstable regions.
    The red square shows parameter regions where the observables computed on the simulated manuscript populations are consistent with observations made for Medieval French epics or with plausible estimates made using different methods.
    }
    \label{fig:phaseDiags}
\end{figure}

As for the remaining free parameters, namely the base ``birth'' or copy rate, $\lambda$, and the ``death'' rate $\mu$, we explored all possible pairs of values of these parameters within their range (from $0$ to $1$ in theory, but reduced here to $10^{-4}$ to $10^{-3}$ by field expertise, i.e.~rough estimates from philological knowledge)\footnote{%
    Historical and philological knowledge of loss rates is very scarce and elusive, 
    but can still be approached from various angles, such as the collection of data from ancient library catalogues, inventories, wills, as well as allusions and intertextuality  \cite{wilson_r_m_lost_1952,bardon_litterature_1952,buringh_medieval_2010}. 
    Buringh \cite{buringh_medieval_2010} provides estimates for the Latin West, with a geometric mean of loss around -25\% per century, with variations from -11\% in the 9th to -32\% in the 14th and 15th centuries (with local variations between medieval institutions from –3\% to –71\% per century). The global loss rate for non-illustrated manuscripts of several well known collections have been estimated around 93-97\% \cite{kestemont_estimating_2020,neddermeyer_von_1998,wijsman_luxury_2010,oostrom_stemmen_2013}. But there is a potential bias in accounting only for well known institutional collections, from which some manuscripts are known to have survived: trying to account for fully lost libraries, Buringh \cite{buringh_medieval_2010} is compelled to revise his estimates higher, to -25\% by century until the 12th, up to -43\% in the 15th. For incunabula, using editions whose original number of print made is known, it is possible to gather loss estimates by counting known surviving exemplars in public or private collections: doing so for Venetian incunabula, Trovato \cite{trovato_everything_2014} finds very variable loss rates according to textual and material typology, from 73\% for the \textit{Decretales} printed on parchment to 99.3\% for more popular chivalrous literature (\textit{Orlando furioso} for instance). This shows the importance both of variation in time and space, and of textual contents and material typology. In some extreme cases, loss can be very close to 100\%, for reasons that may combine the fragility of the document form, lack of consideration for the documents or large scale historical events such as political instability, invasions or major cultural changes; examples are provided by cases as different as the Merowingian royal diplomas on papyrus or the Lombard royal charters \cite{ganz_charters_1990}, the Mayan (pre-colombian) manuscripts 
    or medieval notarial acts \cite{holtz_uberlieferungs-_2001}.
    Production estimates have also been attempted on the basis of the quantity of
    sealing wax acquired by a given producer (a chancellery for instance \cite{bautier_introduction_1978}).
    More founded loss estimates have also been gathered by counting how many of the acts mentioned in imperial or royal registers are kept in original or consigned in the archives of the recipients: this gives a loss rate of originals varying from 80\% (acts from the emperor Charles IV in 1360-1361) to 90\% for the acts from Louis X of France, increasing to 99\% for the judgements rendered by his  Parliament, suggesting here as well a massive effect of typological variation \cite{holtz_uberlieferungs-_2001,canteaut_quantifier_2020}, resulting in very strong biases in the body of  documents available to us. 
    For our simulation needs, if we start from Trovato's estimates (potentially more reliable, because based on editions whose original number of copies is known), we get a survival rate whose order of magnitude is between $0.1$ and $0.01$ (between $10^{-2}$ and $10^{-3}$) in 500 years (2000 steps in our model), in similar ranges as Buringh's $0.75$ in 100 years and those reported by Holtz and Canteaut; from this we can deduce a step loss rate for a given total survival rate. For instance, for 1\% survival rate,  
    $(1-\mu)^{2000} = 0.01$, which simplifies to $\mu = 0.002$. So we retain values of $\mu$ between $10^{-4}$ to $10^{-3}$.
    Given that books could not have been produced order of magnitudes higher or slower than they were destroyed (or we would be either drown in medieval manuscripts or keep none), we explore the same range for $\lambda$. Of course, fixed rates are a limitation, and do not yet account for  substantial variations in time (such as massive extinction events, like, e.g., the fall of the Roman empire, the fire in Alexandria library, the shift from \textit{volumen} to \textit{codex} or from caroline to gothic script, etc.).
}. The space of all possible values is usually called the phase space in physics and other mathematical sciences, and a representation of the value taken by a given observable quantity (eg the extinction probability) when parameters are varied across the phase space is called the phase diagram of this quantity. We thus produced phase diagrams for a number of relevant observables~(Fig.~\ref{fig:phaseDiags}). Note that, since there is a stochastic component in the model, each phase diagram is computed by averaging over the results of a relatively large number of simulations, here $100$ ---~this means that we produced $100$ artificial manuscript traditions for each pair $(\lambda,\mu)\in [10^{-4},10^{-3}]$, varying values by increments of $10^{-4}$; hence each phase diagram required $10~000$~simulations.

The approach then consists in identifying, within the phase diagrams (also called heat maps), regions in which the values for the observables are consistent either with measured quantities (as is done in the natural sciences) or with estimates for these quantities coming from other, independent and altogether different, models. We have circled in red such regions on the phase diagrams in~Fig.~\ref{fig:phaseDiags}. 

These heat maps show that the results obtained through these simulations are internally consistent in terms of not only population size and survival rates, but also in terms of structural properties of the resulting trees. In particular, the results obtained for a ratio $\frac{\mu}{\lambda}$ between $\frac{5}{8}$ and $\frac{6}{7}$ are surprisingly consistent with the observed properties of some medieval traditions, in particular those from chivalric narratives in Old French. In particular, values of 0.55 for the survival of works and 0.05 for the survival of manuscripts  (fig.~\ref{fig:phaseDiags}), \textbf{A} and \textbf{B}, red squared area, bottom-right tile) are identical to those provided by Kestemont et al. for Old French chivalric romances, using unrelated methods from ecodiversity   \cite{kestemont_forgotten_2022}. 
Yet, for what regards specifically Old French epics, known as \textit{chanson de geste} --~a genre predating the later form of the \textit{roman}, and whose circulation and reception considerably differs for a long time~--,  the median final population of 2 and the third quartile of LCA outdegree of 2
(though, median LCA Shannon index is 0.69)\footnote{%
    Data about the traditions of the \textit{chansons de geste} follow Vitale-Brovarone's \cite{vitale-brovarone_diffusion_2006} and Camps'  \cite{camps_`chanson_2016}. Information on the shape of stemmata have been computed based on a restriction to \textit{chansons de geste} and deduplicating of the collection provided by OpenStemmata \cite{camps_open_2021,camps_open_2021b}.
}. This would lead us to revise Kestemont et al. estimates to 0.22 (instead of 0.55) for the specific survival of Old French epics (\textit{chansons de geste}) and 0.01 (instead of 0.5) for the survival of epic manuscripts (a figure closer to that observed by Trovato \cite{trovato_everything_2014} for their later Italian successors). On the other hand, specific values, this time, for (Arthurian and Antique matters) \textit{romans} yield a third quartile of LCA outdegree of 3 more coherent with Kestemont et al. general estimate (median tradition size, according to Martina \cite{martina_produzione_2018}, is 2 --~and mean 4.8~-- for the sole \textit{romans en vers}, similarly to \textit{chansons de geste}, but is expected to be higher for later \textit{romans en prose}).

The situation is, for the moment (and until further data is acquired) consistent yet deserving of further inquiry concerning the distribution of age of surviving manuscripts: 
in the simulation's red-squared area, the median date of surviving manuscripts would be in the 800's step (around 200 years after the original) and the median date of the oldest for each tradition in the 150-200 years range. For \textit{chansons de geste}, the median date of surviving manuscripts would be between 1250 and 1300 \cite{vitale-brovarone_diffusion_2006, camps_`chanson_2016} -- 150 to 200 years after the first documents of the genre itself (the end of the 11th century for the composition of the oldest version of the \textit{Roland}). For \textit{romans en vers}, it is,  like the genre itself, slightly offset in time, with a peak between 1275 and 1325 \cite{martina_produzione_2018}. 


\section{Discussion}




Combining previous inquiries by Weitzman and Cisne with the power of computer simulations and the methodology of statistical physics, we are able to reproduce the evolutionary process that underlies the observable data for, at least, some textual traditions such as those from medieval French epics and romances. The results obtained can even corroborate or refine results obtained by unrelated methodologies, such as those recently published by Kestemont et al. \cite{kestemont_forgotten_2022}, indicating that these relatively simple birth-and-death process have relevancy in philology as well as they have in Evolutionary Biology for instance. This method then provides us a way to account both for population dynamics in time, loss or production estimates, as well as the shape of the stemmata (the phylogenies) of manuscripts, answering the century long Bédier observation \cite{bedier_tradition_1928}, whose lack of solution until now has been  at the core of a lasting schism in philological studies.

The range of further investigations opened by this research is considerably large. Models using individual variable rates of $\lambda$ and $\mu$ could be used to account for phenomenon such as efforts of preservation of old and venerable artefacts, or higher selective values of some copies, or accelerated destruction 
due to small scale (e.g., burnt libraries), larger scale (e.g., the Dissolution of English monasteries, French Wars of Religion,…) or global events (e.g., shift in book types such as from \textit{volumen} to \textit{codex} or caroline to \textit{gothic} scripts, major cultural changes like the Renaissance, …). Modelling should also include the actual variation of the texts, to reflect the introduction of variants (mutations) in some families, and processes of transmission of inherited variants, as well as lateral transmission. 

More generally, once having established this `null model', deviations due to different factors should be explored, such as higher selective values for some mutations (\textit{variants}) or individuals, fluctuations in time and space and the existence of different `ecological niches' (e.g., the Anglo-Norman public versus the readers of Franco-Italian epics), typological variation in books or texts, chocks and bottlenecks, etc. 
The question of the age of surviving manuscripts should also be explored and accounted for, especially in the light of potential variations of $\lambda$ and $\mu$ in time. For instance, the demand and rate of copy for a given text could be expected to be highest shortly after its initial release, when it is most fitted to the taste and fashion of the time, perhaps reinforce itself if the text gets a quick breakthrough, and then decrease with the passing of years. Similarly, the rate of destruction could vary at a global or local level, as some shocks lead to peaks of destruction or canonicalisation and conservation efforts lead to lower rates.  
Last but not least, the model should account for non standard transmission, in particular lateral transmission (contamination), a process not uncommon in textual transmission but that is also encountered in the natural world (e.g., lateral gene transfer).



The generality of the models considered here makes them applicable not only to medieval texts, but to any type of cultural transmission, at least in written form, from manuscript circulation to the elaboration of a canon of works.  
Further investigations should try to verify it on the broadest possible range of cases, starting with Western Medieval and Antique texts, but preferably encompassing cultural productions from very different time periods and continents. 


\bibliography{main}



\end{document}